\documentclass[aps,reprint,groupedaddress,longbibliography]{revtex4-1}
\usepackage{graphicx,epsfig,amsmath,amssymb,color}
\usepackage[normalem]{ulem}

\begin{document}
\newcommand{\kT}[0]{k_{\textrm{B}}T}
\newcommand{\myvec}[1]{\boldsymbol{#1}}
\newcommand{\myten}[1]{\mathsf{#1}}

\title{Current fluctuations in nanopores: the effects of electrostatic and hydrodynamic interactions}

\author{Mira Zorkot}
\author{Ramin Golestanian}
\email{ramin.golestanian@physics.ox.ac.uk}
\author{Douwe Jan Bonthuis}
\email{douwe.bonthuis@physics.ox.ac.uk}
\affiliation{Rudolf Peierls Centre for Theoretical Physics, Oxford University, Oxford, OX13NP, United Kingdom}

\date{\today}

\begin{abstract}
Using nonequilibrium Langevin dynamics simulations of an electrolyte with explicit solvent particles, we investigate the effect of hydrodynamic interactions on the power spectrum of ionic nanopore currents.
At low frequency, we find a power-law dependence of the power spectral density, with an exponent depending on the ion density.
Surprisingly, however, the exponent is not affected by the presence of the neutral solvent particles.
We conclude that hydrodynamic interactions do not affect the shape of the power spectrum in the frequency range studied.
\end{abstract}

\maketitle

Hydrodynamic interactions have a strong influence on the dynamics of Brownian particles suspended in a solvent, 
producing self-organized states, nonlinear dynamics, and synchronization \cite{2011_Golestanian, 2009_McWhirter, 2010_Lee, 2010_Kotar}.
Hydrodynamic interactions between objects decay slowly.
Similar to the electrostatic potential, the strength of the hydrodynamic interactions in bulk is inversely proportional to the distance between the particles \cite{1997_Crocker}.
Moreover, the effects of hydrodynamic interactions are extremely sensitive to geometric confinement.
Density perturbations in a fluid between stationary confining walls give rise to a long-time tail in the velocity autocorrelation function of colloidal particles \cite{1997_Hagen}.
Experiments show that hydrodynamic interactions even become independent of the distance between particles inside small pores \cite{2015_Misiunas}.
These hydrodynamic effects have a pronounced effect on the dynamics of larger molecules, such as DNA, translocating through a nanopore \cite{2008_Fyta, 2013_Laohakunakorn}.
Whereas the effect of hydrodynamic interactions on colloidal particles and polymer dynamics has attracted a lot of attention over the past decades, the effect of hydrodynamic interactions on ion dynamics remains largely unexplored.

The combination of experimental measurement and molecular modeling of the power spectral density constitutes a promising technique to study ion motion in unprecedented detail \cite{2008_Fulinski}.
For example, the power spectrum can be used to study the microscopic properties of nanofluidic systems, such as the adsorption of molecules on the walls of a nanometer-scale cavity \cite{2011_Singh-Lemay}.
Recently, we showed that ion correlations at high particle density produce a power-law spectrum at low frequency, with an exponent depending on the ion density \cite{2016_Zorkot}.
Experiments show that the power spectrum of the ionic nanopore current $S\left(\omega\right)$, with $\omega = 2 \pi f$ being the frequency, typically follows a power law $S\left(\omega\right) \propto 1/\omega^{\alpha}$ with $\alpha \approx 1$, 
which is referred to as pink noise, or $1/f$ noise \cite{1988_Weissman_RevModPhys, 2015_Heerema, 2008_Smeets_PNAS}.
The appearance of pink noise in nanopore current measurements is ubiquitous; it is found in a variety of systems, from protein channels and flexible synthetic pores \cite{2000_Bezrukov_PRL, 1997_Wohnsland, 2002_Siwy-Fulinski_PRL}, to solid-state conical pores \cite{2009_Powell-Siwy_PRL}.
The molecular origin of the low-frequency pink noise has been debated for decades \cite{1970_Hooge, 2009_Hoogerheide, 2010_Tasserit}.
However, theoretical analysis of the power spectrum including the multi-body interactions between the ions and the effect of hydrodynamics remains challenging.
Therefore, although hydrodynamic interactions are usually present in experimental studies, their effects on the frequency dependence of the power spectral density are unknown.
The situation has changed with the recent advance of fast and versatile molecular simulation techniques, which now allow a systematic computational investigation.

In this manuscript, we present a Langevin dynamics simulation study of the ionic current through a nanometer-scale pore filled with an electrolyte, using the Espresso molecular dynamics package \cite{2006_Limbach}.
The electrolyte is modeled by ions in an explicit solvent.
For the solvent, we use a coarse-grained description of neutral, nonpolar Lennard-Jones particles.
To systematically study the effect of hydrodynamic interactions, we vary the density of both the ions and the solvent particles independently.
We calculate the power spectral density of the ion current and compare the results with simulations without solvent and with a linearized mean-field theory of ion currents without hydrodynamic interactions.
Whereas an increase in the ion density directly causes a power-law behavior of the power spectrum at low frequency, introducing hydrodynamic interactions by increasing the solvent density does not have the same effect.

\begin{figure}
\includegraphics[width=0.9\columnwidth]{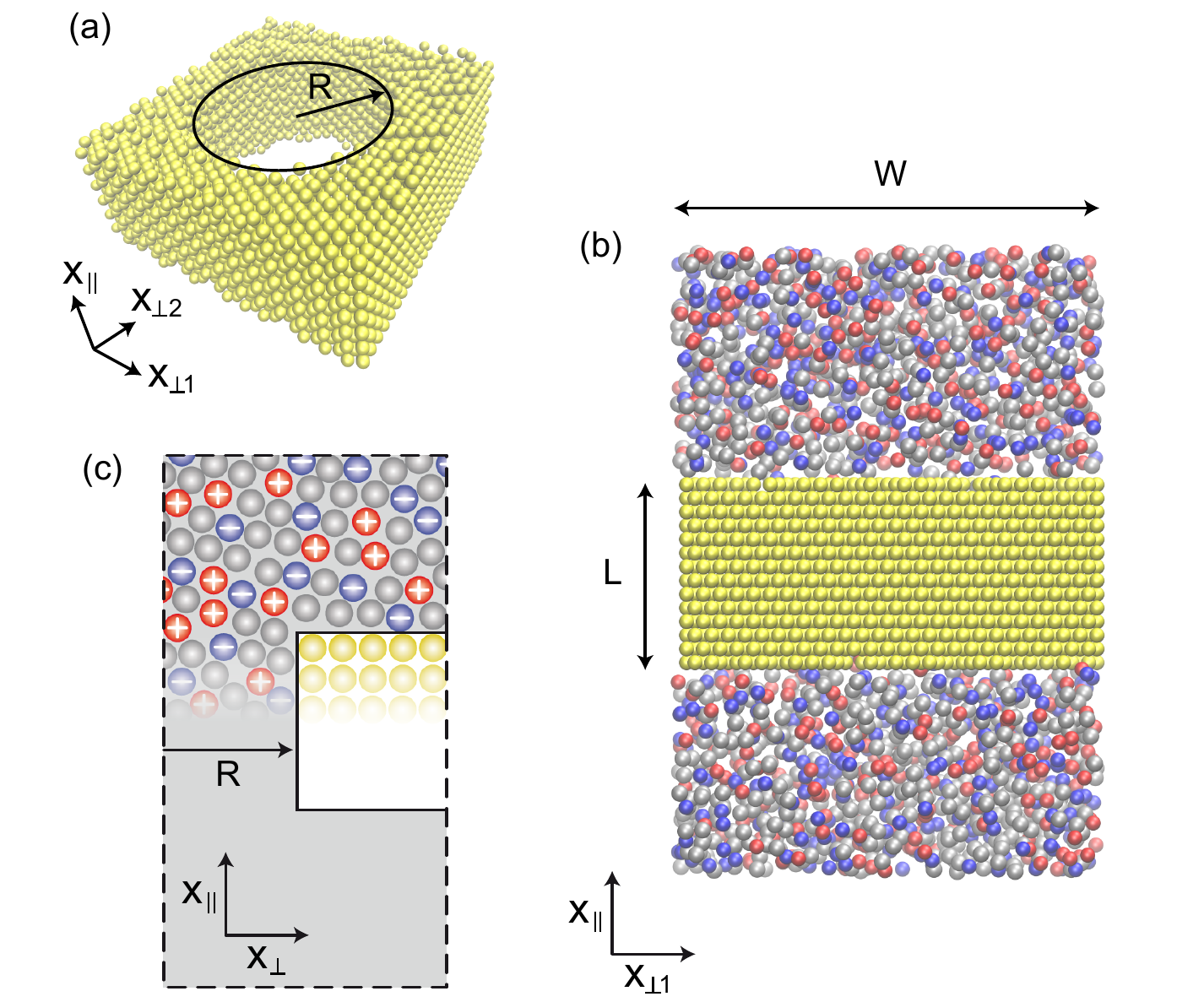}
\caption{
(a) Simulation snapshot of the membrane with the pore of radius $R$. Ions and solvent are not shown.
(b) Side view of the simulation box with ions shown in blue and red, and solvent particles in grey. 
(c) Schematic drawing of the nanopore, with either $x_{\perp}^2 = x_{\perp1}^2+x_{\perp2}^2$ (cylindrical coordinates, Eq. \ref{eqn:area-cylinder}) or $x_{\perp} = x_{\perp1} = x_{\perp2}$ (cartesian coordinates, Eq. \ref{eqn:area-square}).
}
\label{fig:snapshot}
\end{figure}

\subsubsection{Simulation model}

We use the simulation package Espresso \cite{2006_Limbach} to set up Langevin dynamics simulations of a nanopore filled with a mixture of monovalent positive and negative ions and neutral solvent particles (Fig. \ref{fig:snapshot}).
The Langevin equation for particle $i$ is expressed as
\begin{equation} \label{eqn:langevin}
m_{i} \frac{\partial \myvec{u}_{i}}{\partial t} = -\sum_{j\neq i} \nabla V_{ij} + \myvec{F}_{i} - {\gamma} \myvec{u}_{i} + \myvec{\xi}_{i},
\end{equation}
where $\myvec{\xi}_{i}\left(t\right)$ is the stochastic force satisfying $\langle \myvec{\xi}_{\textit{i}}\left(t\right) \cdot \myvec{\xi}_{\textit{j}}\left(t'\right)\rangle = 6 k_{B} T \gamma \delta_{\textit{ij}} \delta\left(t-t'\right)$,
$\myvec{u}_i$ and $m_i = 1$~$k_{B}T\tau^2/$\AA$^2$~denote the velocity and the mass, respectively, and $\myvec{F}_i$ is an external force applied to the particle.
The thermal energy equals $k_BT$ and we use $\gamma = 1$~$k_BT\tau$/\AA$^2$.
By using an equal and arbitrary mass for all particles, $m_i$ is incorporated in the time scale $\tau$. 
Short-ranged interactions between pairs of particles
are modeled by a Weeks-Chandler-Andersen (shifted Lennard-Jones) potential $V_{ij}\left(r_{ij}\right)$,
\begin{equation} \label{eqn:potential}
V_{ij} = l_{B}k_{B}T\frac{Q_{i}Q_{j}}{r_{ij}} + 4 \epsilon_{ij} \left[ \left(\frac{\sigma_{ij}}{r_{ij}}\right)^{12}-\left(\frac{\sigma_{ij}}{r_{ij}}\right)^{6}\right] + \epsilon_{ij},
\end{equation}
where $\sigma_{ij}$ denotes the distance between particles $i$ and $j$, $Q_i$ is the charge of particle $i$ in units of the elementary charge $e$, and $\epsilon_{ij}$ represents the interaction strength. 
The Bjerrum length $l_{B} = e^2/\left(4\pi\varepsilon\varepsilon_0 k_B T\right)$, and the distance between any pair of particles is given by $r_{ij}$.
The Lennard-Jones interaction is truncated at $r_{ij}=2^{\frac{1}{6}}\sigma_{ij}$ for all combinations of $i$,$j$.
The electric field on the ions is represented by a force applied inside the pore in the $x_{\parallel}$ direction, which is the direction along the length $L$ of the pore (Fig. \ref{fig:snapshot}),
\begin{equation} \label{eq:external-force}
\myvec{F}_{i} = \Bigg\{ {\begin{array}{l l} 
Q_{i} E_{\parallel} & \quad \text{if $0<x_{\parallel}<L$ }\\
 0 & \quad \text{otherwise}.
  \end{array}}
\end{equation}
The electric field is varied between $E_{\parallel}$=0.3 and 1.6~$k_{B}T/\left(e\text{\AA}\right)$.

The simulations are performed in a cylindrical nanopore with a radius ranging from $19$~\AA~ to $30$~\AA, permeating a rigid membrane with a width of $W = 96$~\AA~and a length $L = 48$~\AA~(Fig. \ref{fig:snapshot}).
We use an increasing solvent density of $C_s = \{5.5\cdot 10^{-4},2.1\cdot 10^{-3},5.5\cdot 10^{-3}\}$~\AA$^{-3}$,
which are all well below the bulk freezing density of the Weeks-Chandler-Andersen fluid model.
Compared to the molecular density of water, the maximum density corresponds to a coarse-grained force field where each particle represents approximately 6 water molecules.
A smooth surface induces a crystalline order in the fluid, over a range depending on the molecular properties of the liquid \cite{2012_Hadley}. 
For a fluid consisting of identical Lennard-Jones spheres, the induced order propagates over a distance larger than our simulation box.
Therefore, to prevent crystallization of the Lennard-Jones fluid, we perturb the uniform membrane surface by randomly removing half of the particles from the outer layer.
The membrane particles are frozen, and for the membrane-ion, membrane-solvent, ion-solvent, ion-ion and solvent-solvent interactions we use $\epsilon_{ij} = 2$~$k_BT$ and $\sigma_{ij} = 4.7$~\AA.
For the membrane and solvent particles we use $Q_i = 0$, and for the ions we use $Q_i = \pm 1$.
The ion concentration is varied according to $C_i = \{5.5\cdot 10^{-5},5.5\cdot 10^{-4},1.0\cdot 10^{-3},2.1\cdot 10^{-3}\}$~\AA$^{-3}$.

\begin{figure}
\includegraphics[width=0.9\columnwidth]{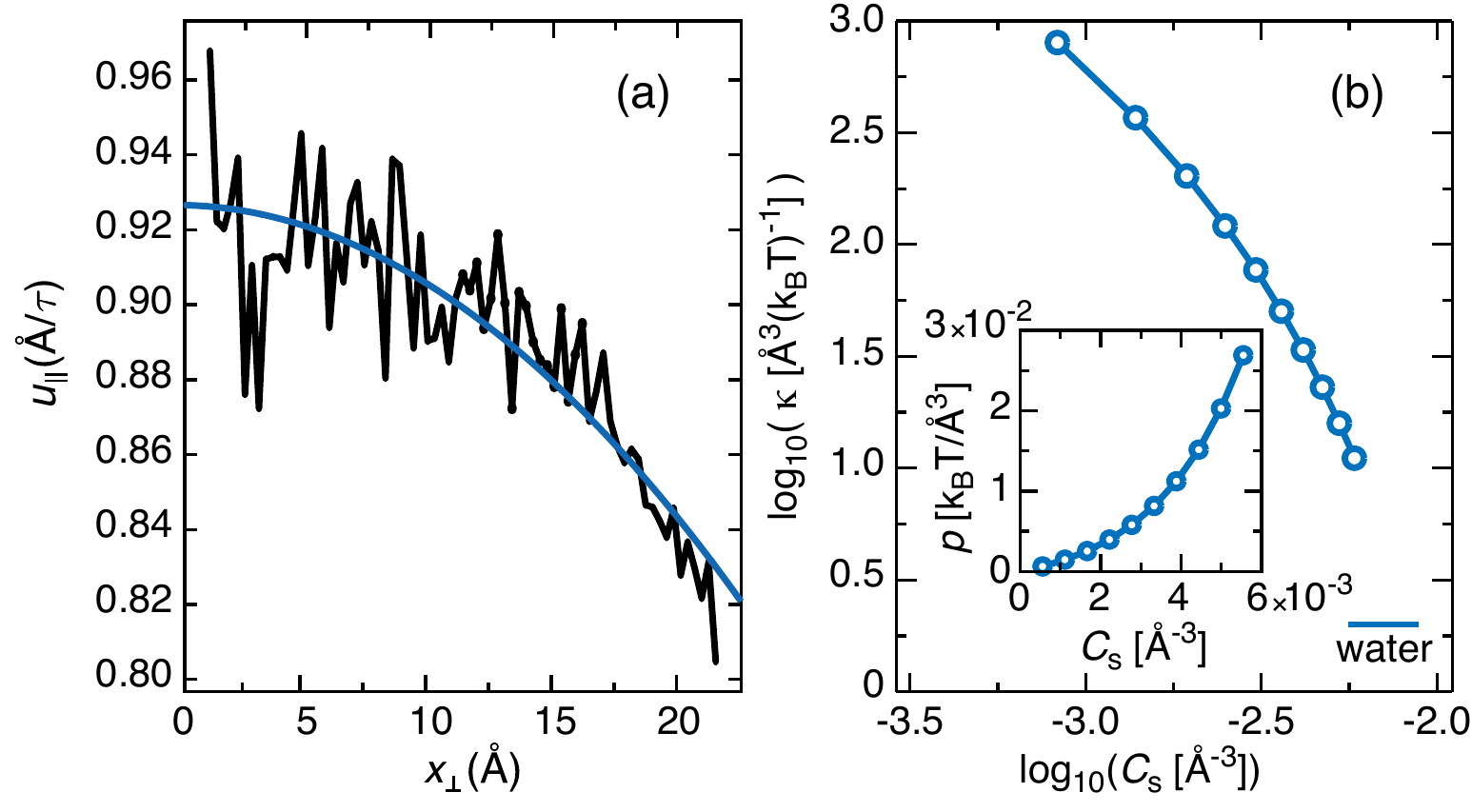}
\caption{
(a) The connected black dots represent the tangential flow velocity $u_{\parallel}$ as a function of the radial position $x_{\perp}$ in response to an applied pressure gradient, and the solid blue line is the fitted Poiseuille flow profile (Eq. \ref{eqn:poiseuille}).
(b) The bulk compressibility $\kappa$ as a function of the solvent concentration $C_s$.
The experimental compressibility of water at room temperature is shown as a reference.
The inset shows the pressure $p$ as a function of $C_s$.
}
\label{fig:poiseuille}
\end{figure}

When the motion of particle $i$ perturbs the surrounding solvent, the hydrodynamic signal diffuses at a rate governed by the kinematic viscosity $\nu$.
For hydrodynamic interactions to occur, this viscous momentum must diffuse much faster than the particle itself.
The relation is governed by the Schmidt number $\textit{Sc} = \nu/D$, with $D$ being the diffusion coefficient of the solvent particles.
To verify that the coarse-grained solvent particles produce hydrodynamic interactions in the strongly confined environment of the nanopore, we simulate a pressure difference across the length of the channel by applying a constant force to all particles inside a pore filled with pure solvent.
We calculate the fluid velocity as a function of the radial coordinate, averaged across the length of the channel.
The flow of ions in a slit-like cylindrical channel forms a Hagen-Poiseuille flow profile with a finite slip length $b$,
\begin{equation} \label{eqn:poiseuille}
u_{\parallel}=-\dfrac{F_{\parallel}}{4 m_i \nu}\left[R^2+2bR-x_{\perp}^2\right],
\end{equation}
with $R$ being the radius of the pore, ${F_{\parallel} C_s = \nabla_{\parallel} p}$ being the pressure gradient across the length of the pore, which in our case is derived from the uniform applied force $F_{\parallel} = 0.8$~$k_BT$/\AA~on the solvent particles inside the pore, which have mass $m_i = 1$~$k_{B}T\tau^2/$\AA$^2$ and number density $C_s$.
We show the velocity profile in Fig. \ref{fig:poiseuille}(a) for the lowest solvent density $C_s = 5.5\cdot 10^{-4}$~\AA$^{-3}$.
The fit of Eq. \ref{eqn:poiseuille} yields $\nu = 900$~\AA$^2$/$\tau$, which in combination with $D = 1$~\AA$^2$/$\tau$ yields $\textit{Sc} = 900$.
As the Schmidt number at higher solvent densities is even higher, all our simulations satisfy the conditions for hydrodynamic interactions.

Apart from propagation by viscous momentum diffusion, hydrodynamic interactions are transmitted by sound wave propagation.
In an incompressible fluid, the sound velocity is infinite, and the viscous momentum diffusion is solely responsible for the time evolution of the hydrodynamic interactions.
The compressibility of our model solvent is finite, however, depending on the solvent density $C_s$, which might have implications for the hydrodynamic interactions \cite{2006_Padding-Louis}.
We calculate the isothermal compressibility from the pressure $p$ as a function of solvent density $C_s$ in separate bulk simulations using $\kappa_T = (\textrm{d}\ln(C_s)/\textrm{d}p)_T$, with $p$ being the pressure, see Fig. \ref{fig:poiseuille}(b).
The compressibility is varied over two orders of magnitude as we change the solvent density.
Nevertheless,
the compressibility of water, equal to $\kappa_T = 2$~\AA$^3$/($k_BT$), is still a factor $5$ below the compressibility of our highest-density solution.
We quantify the effect of the compressibility by calculating the sound velocity $u_{s} = \sqrt{\gamma/\left(m_i C_{s} \kappa_T\right)}$, with $\gamma$ being the heat capacity ratio, which is of order $\gamma \sim 1$ in a liquid.
The sound velocity increases drastically when we change the solvent density in our simulations, from $u_{s} = 3$~\AA/$\tau$ at the lowest density to $u_{s} = 239$~\AA/$\tau$ at the highest density. 
The importance of the compressibility effects is estimated from the Mach number $\textit{Ma} = \sqrt{k_BT/m_i}/u_{s}$, where
the estimate of the thermal velocity $\sqrt{k_B T/m_i}$ is being used as the typical velocity of the particles.
As $\textit{Ma}$ is well below $1$ in all our simulations, the compressibility is not expected to have a large effect on the hydrodynamic properties \cite{2006_Padding-Louis}.

The stochastic force $\myvec{\xi}$ in the Langevin dynamics simulations provides a truncation length beyond which $\myvec{\xi}$ exceeds the force due to hydrodynamic interactions \cite{2006_Padding-Louis}.
Quantification is complicated, because the truncation length depends on the magnitude of the force from which the hydrodynamic interactions originate.
As the interparticle forces in the system reach very high values, however, a part of the long-ranged hydrodynamic interactions will be preserved.

\subsubsection{Linearized mean-field theory}

We derive a theoretical description of the noise spectrum of the ionic current following our previous analysis \cite{2016_Zorkot}.
The expression for the power spectral density $S\left(\omega\right)$ is derived for monovalent ions in implicit water.
Therefore, the following derivation does not include the effect of hydrodynamic interactions.
Comparison with the simulation results allows us to study the effect that the hydrodynamic interactions in the simulations have on the power spectral density.
Ion-ion correlations, which are responsible for the low-frequency power-law increase of the power spectrum at high ion density \cite{2016_Zorkot}, are also absent from this theoretical model.
We consider a system consisting of a cylindrical nanopore of length $L$ and radius $R$ connecting two reservoirs (Fig. \ref{fig:snapshot}), and calculate the flux density $\myvec{J}^{\pm}\left(\myvec{x},t\right)$ of positive and negative ions inside the nanopore,
with $\myvec{x}$ denoting the position in three dimensions and $t$ denoting the time.
The ion concentrations $C^{\pm}\left(\myvec{x},t\right)$ are governed by the continuity equation,
\begin{equation} \label{eqn:fick}
\frac{\partial C^{\pm}}{\partial t} + \nabla \cdot \myvec{J}^{\pm} = 0.
\end{equation}
The corresponding flux densities $\myvec{J}^{\pm}\left(\myvec{x},t\right)$ are given by the Nernst-Planck equation,
\begin{equation} \label{eqn:nernst-planck}
\myvec{J}^{\pm} = -D^{\pm}  \nabla C^{\pm} \mp D^{\pm}C^{\pm} \frac{ e \myvec{E}}{k_{B}T} + \myvec{\eta}^{\pm},
\end{equation}
where $\myvec{E}\left(\myvec{x},t\right)$ is the applied electric field, $e$ denotes the elementary charge, 
and $\myvec{\eta}^{\pm}\left(\myvec{x},t\right)$ denotes the thermal noise that accounts for fluctuations in the environment; most importantly the effect of the implicit water on the ions dynamics. 
From here, we switch to index notation where $\alpha$, $\beta$, and $\gamma$ correspond to the three components of our coordinate system.
To simplify the notation, we assume $D^+ = D^- = D$ and $\myvec{\eta}^+ = \myvec{\eta}^- = \myvec{\eta}$.
After applying a standard Fourier transform to Eqs. \ref{eqn:fick} and \ref{eqn:nernst-planck}, we find
\begin{equation} \label{eq:er8}
\widetilde{J}^{\pm}_{\alpha} = \sum^{3}_{\beta=1} \left[ -\frac{iD}{\omega} \, q_{\alpha} q_{\beta} \widetilde{J}^{\pm}_{\beta} \mp \frac{D\,eE_{\alpha}}{\omega\, k_BT} \, q_{\beta} \widetilde{J}^{\pm}_{\beta} - \widetilde{\eta}_{\alpha} \right],
\end{equation}
with $\widetilde{...}$ denoting the Fourier transform, $\myvec{q}$ being the wave vector, and $\omega$ being the frequency. 
Rewriting Eq. \ref{eq:er8} leads to
\begin{equation} \label{eq:er10}
\widetilde{\eta}_{\alpha} = -\sum^{3}_{\beta=1} \widetilde{J}^{\pm}_{\beta} M_{\alpha\beta},
\end{equation}
where $M_{\alpha\beta}$ denotes the matrix
\begin{equation} \label{eq:er11}
M_{\alpha\beta}=\delta_{\alpha \beta} + \frac{iD}{\omega} \, q_{\alpha}q_{\beta} \pm \frac{D\,eE_{\alpha}}{\omega \, k_BT} \, q_{\beta}.
\end{equation}
Combining Eqs. \ref{eq:er10} and \ref{eq:er11} and solving for $\widetilde{J}^{\pm}_{\gamma}$, we find
\begin{equation}\label{eq:er13}
\begin{split}
\widetilde{J}^{\pm}_{\gamma} &= \frac{1}{\mathrm{det}(M)}\left[\sum^{3}_{\beta=1} \left[\frac{iD}{\omega}q_{\gamma}{q_{\beta}\widetilde{\eta}_{\beta}} \pm \frac{D\,eE_{\gamma}}{\omega\,k_BT} \, q_{\beta} \widetilde{\eta}_{\beta} \right] \right. \\
 &\qquad - \left. \widetilde{\eta}_{\gamma}\sum^{3}_{\beta=1}{\left[\frac{1}{3} + \frac{iD}{\omega}q_{\beta}{q_{\beta}}
\pm \frac{D\,e{E}_{\beta}}{\omega \, k_BT} \, q_{\beta}\right]} \right],
\end{split}
\end{equation}
with $\mathrm{det}(M)$ denoting the determinant of $M$.
Within the geometry of the pore, there is one parallel ($\parallel$) direction, and two equivalent perpendicular ($\perp$1, $\perp$2) directions, see Fig. \ref{fig:snapshot}(a).
The electric-field is nonzero only in parallel direction $\myvec{E} = (0,0,E_{\parallel})$.
Therefore, the flux in the parallel direction becomes
\begin{equation} \label{eq:er14}
\begin{split}
\widetilde{J}^{\pm}_{\parallel}(\myvec{q},\omega) &=\frac{ \frac{iD}{\omega} \big[{q_{\perp 1}q_{\parallel}\widetilde{\eta}_{\perp 1}} + {q_{\perp 2}q_{\parallel} \widetilde{\eta}_{\perp 2}} + {q_{\perp 1}^{2} \widetilde{\eta}_{\parallel}} + {q_{\perp 2}^{2} \widetilde{\eta}_{\parallel}}\big] }{1 + \frac{iD}{\omega} \big[q_{\parallel}^2 + q_{\perp 1}^2 + q_{\perp 2}^2 \big] \pm \frac{D\,eE_{\parallel}}{\omega\,k_BT} q_{\parallel}}\\
&\qquad + \frac{ \mp\frac{D\,eE_{\parallel}}{\omega\,k_BT}\big[{q_{\perp 1} \widetilde{\eta}_{\perp 1}} + {q_{\perp 2} \widetilde{\eta}_{\perp 2}}\big] + \widetilde{\eta}_{\parallel}}{1 + \frac{iD}{\omega}\big[q_{\parallel}^2 + q_{\perp 1}^2 + q_{\perp 2}^2\big] \pm \frac{D\,e{E}_{\parallel}}{\omega\,k_BT}q_{\parallel}},
\end{split}
\end{equation}
with $q_{\perp 1}$ and $q_{\perp 2}$ being the two independent wave vectors in the plane of the membrane.
As the random force is applied to every individual particle, the power spectrum of the thermal noise in our implicit-solvent model is proportional to the ion concentration inside the pore,
\begin{equation} \label{eq:er17}
\begin{split}
\langle \eta_{\gamma}\left(\myvec{x},t\right)\eta_{\beta}(\myvec{x}',t')\rangle  &=  2D C_{V} \delta_{\gamma\beta}\delta(\myvec{x}-\myvec{x}')\delta(t-t') \\
\langle \widetilde{\eta}_{\gamma}(\myvec{q},\omega)\widetilde{\eta}_{\beta}(\myvec{q}',\omega')\rangle &= 2D C_{V} \delta_{\gamma\beta}(2\pi)^{4}\delta(\myvec{q}+\myvec{q}') \delta(\omega+\omega'),
\end{split}
\end{equation}
with $C_{V} = \langle N\rangle/(\pi R^2 L)$ being the average number of ions per unit volume in the pore, which is proportional to the bulk ion concentration $C_{i}$, but depends nontrivially on the radius $R$, the length $L$, the electric field, and the interionic interaction potential.
Introducing short-hand notation, we derive from Eqs. \ref{eq:er14}-\ref{eq:er17}
\begin{equation} \label{eqn:current}
\begin{split}
\langle| & \widetilde{J}_{\parallel}^{+} - \widetilde{J}_{\parallel}^{-}|^2\rangle \equiv \int\!\!\int\!\!\int \frac{\textrm{d}q_{\perp1}'\textrm{d}q_{\perp2}'\textrm{d}q_{\parallel}'}{(2\pi)^3}\int \frac{\textrm{d}\omega'}{2\pi} \Big[\\
\langle\big[ &\widetilde{J}^{+}_{\parallel}(\myvec{q},\omega)-\widetilde{J}^{-}_{\parallel}(\myvec{q},\omega)\big]
       \big[\widetilde{J}^{+}_{\parallel}(\myvec{q}',\omega')-\widetilde{J}^{-}_{\parallel}(\myvec{q}',\omega')\big]\rangle \Big] \\
& = \frac{8 D C_{V} \Big[\frac{eE_{\parallel}}{kT}\Big]^2 \Big[\frac{\omega^2}{D^2} + \left(q_{\perp1}^{2}+q_{\perp2}^{2}\right)^{2}\Big] \, \myvec{q}^{2}}
       {\Big(\left(\myvec{q}^{2}\right)^2  \!\! + \Big[\frac{eE_{\parallel} }{kT}\Big]^2 \! q_{\parallel}^2 \! 
   - \frac{\omega^2}{D^2} \! \Big)^2 \!\!  + 4 \frac{\omega^2}{D^2} \left(\myvec{q}^{2}\right)^2},
\end{split}
\end{equation}
with $\myvec{q}^2 = q_{\perp1}^{2} + q_{\perp2}^{2} + q_{\parallel}^{2}$.

The two-sided power spectral density $S\left(\omega\right)$ of the current $I_{\parallel}\left(t\right)$ defined on the domain $0 < t <T$ is given by the limit of $T \to \infty$ of
\begin{equation}
S\left(\omega\right) = \frac{1}{T} \langle | \tilde{I}_{\parallel}\left(\omega\right) |^2 \rangle = \frac{1}{T} \langle \tilde{I}_{\parallel} \left(\omega\right) \tilde{I}_{\parallel}\left(-\omega\right) \rangle,
\end{equation}
which can be written as
\begin{equation}
S\left(\omega\right) = \frac{1}{T} \int_T \!\textrm{d}t \int_T \!\textrm{d}t' \, e^{-i\omega\left(t-t'\right)} \, \langle I_{\parallel}\left(t\right) I_{\parallel}\left(t'\right) \rangle.
\end{equation}
We rewrite $I_{\parallel}\left(t\right)$ as the integral of the current density $J^+_{\parallel}\left(\myvec{x},t\right)-J^-_{\parallel}\left(\myvec{x},t\right)$ at a given position in the direction of $x_{\parallel}$
over the lateral surface area $A$ of the pore,
\begin{equation} \label{eqn:S_of_J}
\begin{split}
& S\left(\omega\right) = \frac{1}{T} \int_T \!\!\textrm{d}t \! \int_T \!\!\textrm{d}t' 
\, e^{-i\omega\left(t-t'\right)} \\
& \! \int\!\!\!\int_A \!\!\textrm{d}x_{\perp1}  \textrm{d}x_{\perp2} \! \int \!\!\textrm{d}x_{\parallel}
  \! \int\!\!\!\int_A \!\!\textrm{d}x_{\perp1}' \textrm{d}x_{\perp2}'\! \int \!\!\textrm{d}x_{\parallel}' \delta(x_{\parallel})\delta(x_{\parallel}')\\
&\langle 
\big[J^+_{\parallel}\left(\myvec{x},t\right) - J^-_{\parallel}\left(\myvec{x},t\right)\big]
\big[J^+_{\parallel}\left(\myvec{x}',t'\right) - J^-_{\parallel}\left(\myvec{x}',t'\right)\big]
\rangle.
\end{split}
\end{equation}
Some mathematical manipulation yields in the limit $T\rightarrow \infty$,
\begin{equation} \label{eqn:psd}
\begin{split}
S\left(\omega\right) &= \int \! \frac{\textrm{d}q_{\perp1} }{2\pi} \! \int \! \frac{\textrm{d}q_{\perp2} }{2\pi} \! \int \! \frac{\textrm{d}q_{\parallel} }{2\pi} \widetilde{A}^2(q_{\perp}) \\
& \quad\quad\quad \langle |\big[\tilde{J}^+_{\parallel}\left( \myvec{q}, \omega\right) - \tilde{J}^-_{\parallel}\left( \myvec{q}, \omega\right)\big] |^2 \rangle\\
& =  8 \! \int \!\! \int^{\pi\varLambda^{-1}}_{\pi R^{-1}} \! \frac{\textrm{d} q_{\perp1} \textrm{d} q_{\perp2}}{(2\pi)^2} 
    \!  \int^{\pi\varLambda^{-1}}_{\pi L^{-1}} \! \frac{\textrm{d} q_{\parallel}}{2\pi} \, \widetilde{A}^2 \, \langle |\widetilde{J}_{\parallel}^{+}-\widetilde{J}_{\parallel}^{-}|^2\rangle,
\end{split}
\end{equation}
with $\varLambda$ being the small-scale cut-off length, introduced because of the finite particle size.
The Fourier-transformed area function in Eq. \ref{eqn:psd} is given by
\begin{equation} \label{eqn:area-derivation}
\widetilde{A}(q_{\perp1},q_{\perp2})  =
  \! \int\!\!\!\int_A \!\!\textrm{d}x_{\perp1}  \textrm{d}x_{\perp2} \,  e^{-i(q_{\perp1}x_{\perp1}+q_{\perp2}x_{\perp2})}.
\end{equation}
The integral in Eq. \ref{eqn:area-derivation} is performed over the lateral area $A$ of the pore, which is approximately circular.
However, because our cylindrical direct space does not map exactly to a cylindrical reciprocal space,
we use two different approximations to calculate the integral (Fig .\ref{fig:snapshot}(c)).
First,
integrating over a square of sides $2R$ gives
\begin{equation} \label{eqn:area-square}
\widetilde{A}(q_{\perp1},q_{\perp2}) = \frac{4 \sin\left[q_{\perp1}R\right]\sin\left[q_{\perp2}R\right]}{q_{\perp1}q_{\perp2}}.
\end{equation}
Alternatively,
integrating over a circle of radius $R$ gives
\begin{equation} \label{eqn:area-cylinder}
\begin{split}
\widetilde{A}\left(q_{\perp}\right) &= \int_{0}^{2\pi}\!\!\!\textrm{d}\theta \! \int_{0}^{R}\!\!\! \textrm{d}x_{\perp} \, 
x_{\perp} \,  e^{-iq_{\perp}x_{\perp}\cos\left(\phi - \theta\right)} \\
& = 2\pi R^2 \; \frac{J_{1}\left[q_{\perp} R\right]}{q_{\perp} R}.
\end{split}
\end{equation}
with $x_{\perp} = \sqrt{x_{\perp1}^2 + x_{\perp2}^2}$ and $\theta = \arctan(x_{\perp2}/x_{\perp1})$ being the cylindrical coordinates, $q_{\perp} = \sqrt{q_{\perp1}^2+q_{\perp2}^2}$ and $\phi = \arctan(q_{\perp2}/q_{\perp1})$ being the polar coordinates in reciprocal space, and $J_{1}$ being the first order Bessel function of the first kind.
The primary difference between Eqs. \ref{eqn:area-square} and \ref{eqn:area-cylinder} is the amplitude of the calculated noise spectrum \cite{2016_Zorkot}.
Contrary to the circular area, however, the square area can be mapped directly to reciprocal space, enabling a straightforward evaluation of Eq. \ref{eqn:psd}.
Therefore, we use Eq. \ref{eqn:area-square} for all the curves in the present paper.
Together with Eqs. \ref{eqn:current} and \ref{eqn:area-square}, Eq. \ref{eqn:psd} is solved numerically to get the linearized mean-field prediction of $S(\omega)$.

\subsubsection{Results and discussion}

\begin{figure}[tb]
\includegraphics[width=0.9\columnwidth]{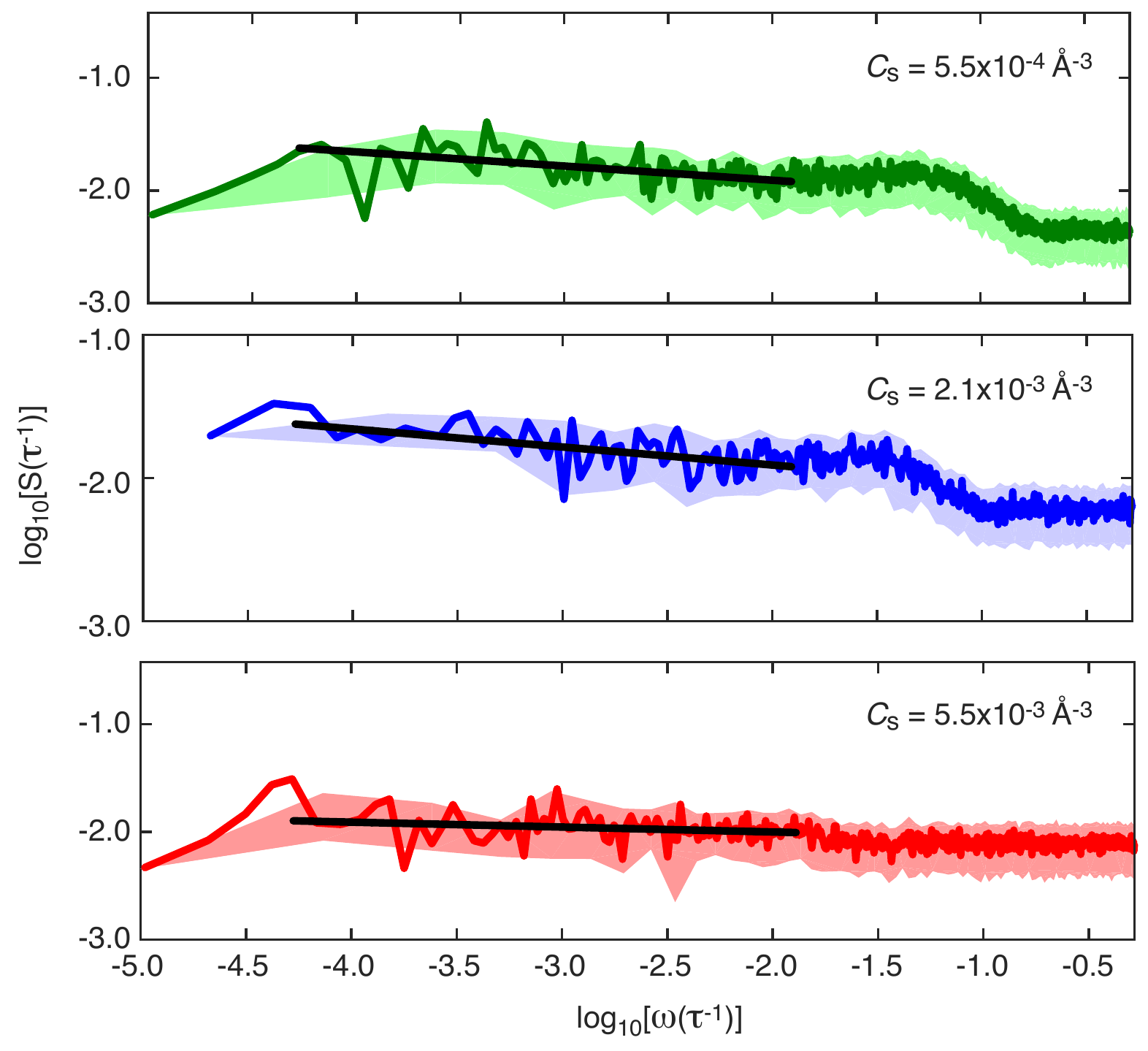}
\caption{
The power spectra of the ion current through pores with different solvent densities $C_s$ in units of the inverse time scale $\tau^{-1}$.
Solid black lines indicate the fit with $S \sim 1/\omega^{a}$.
We use a fixed ion concentration {$C_i = 5.5 \times 10^{-4}$~\AA$^{-3}$}, radius $R=25$~\AA~and an applied electric field $E_{\parallel} = 1.6$~$k_B T/(e\text{\AA})$.
}
\label{fig:solvent-concentration}
\end{figure}

We calculate the power spectral density of the ion current in simulations with three different solvent densities $C_s$ (Fig. \ref{fig:solvent-concentration}).
At low solvent density, the curves exhibit a transition around $\omega = 0.1/\tau$, similar to the implicit solvent case \cite{2016_Zorkot}.
With increasing solvent density, the transition becomes less pronounced due to an increase in the high-frequency noise level.
Surprisingly, the increasing solvent density does not induce any alteration of the power spectral density at low frequency, even for a tenfold increase in solvent density.

\begin{figure}[tb]
\includegraphics[width=0.9\columnwidth]{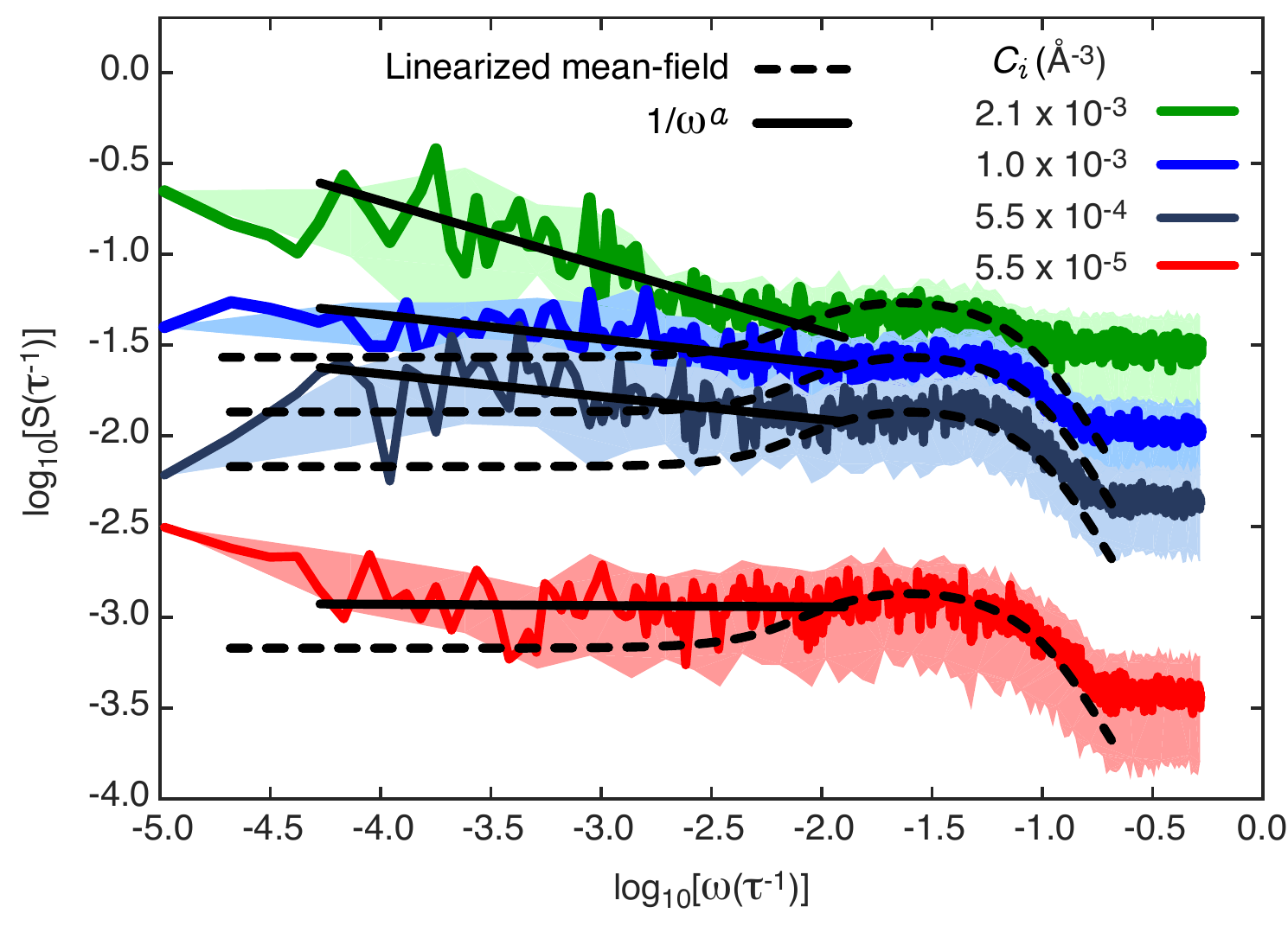}
\caption{
The power spectral density $S\left(\omega\right)$ of the ionic current in units of the inverse of the time scale $\tau$, as a function of the frequency $\omega$ on a log-log scale, 
for different ion concentration $C_i$, in combination with a solvent density of $C_s=5.5\times 10^{-4}$~\AA$^{-3}$.
The solid colored lines show the simulation results and the shaded region represents the standard deviation that we get when applying a block-averaging method to improve the readability.
The dashed lines represent the fits derived from the linearized mean field theory (Eqs. \ref{eqn:current}, \ref{eqn:psd} and \ref{eqn:area-square}),
with parameters taken from the simulations: applied electric field $E_{\parallel}=1.6$~$k_BT$/($e$\AA), pore radius $R=25$~\AA, diffusion coefficient $D = 1$~\AA$^2/\tau$, and the small-scale cutoff length is set to a value of the order of the ion size, $\Lambda = 2.5$~\AA, for all curves.
The solid black lines indicate the fit with $S \sim 1/\omega^{a}$.
}
\label{fig:ion-concentration}
\end{figure}

We verify the effect of increasing ion concentration on the power spectral density in the presence of hydrodynamic interactions (Fig. \ref{fig:ion-concentration}).
The amplitude of the noise increases with increasing ion concentration, and the transition frequency shifts to slightly lower values.
Most strikingly, however, is the change of the behavior at low frequency.
The power spectral density exhibits a power law, with an exponent that increases sharply with increasing ion concentration.
These results are similar to the results found in simulations with implicit solvent \cite{2016_Zorkot}.
However, as the curves in Fig. \ref{fig:ion-concentration} extend to higher ion concentration then treated previously, the new results show that the increase in the exponent of the power law continues, reaching $a=0.4$ at an ion concentration of $C_i = 2\cdot10^{-3}$~\AA$^{-3}$.

To test the effect of the hydrodynamic interactions, we fit the linearized mean field theory, which does not take hydrodynamic interactions into account, to the curves in Fig. \ref{fig:ion-concentration}.
Apart from the low-frequency power-law dependence, which is caused by ion-ion correlations \cite{2016_Zorkot}, the simulated curves are well described by the implicit-solvent model.
Remarkably, it is not necessary to take hydrodynamic interactions into account to describe the power spectrum of the ionic current through an electrolyte-filled pore.

\begin{figure}[tb]
\includegraphics[width=0.9\columnwidth]{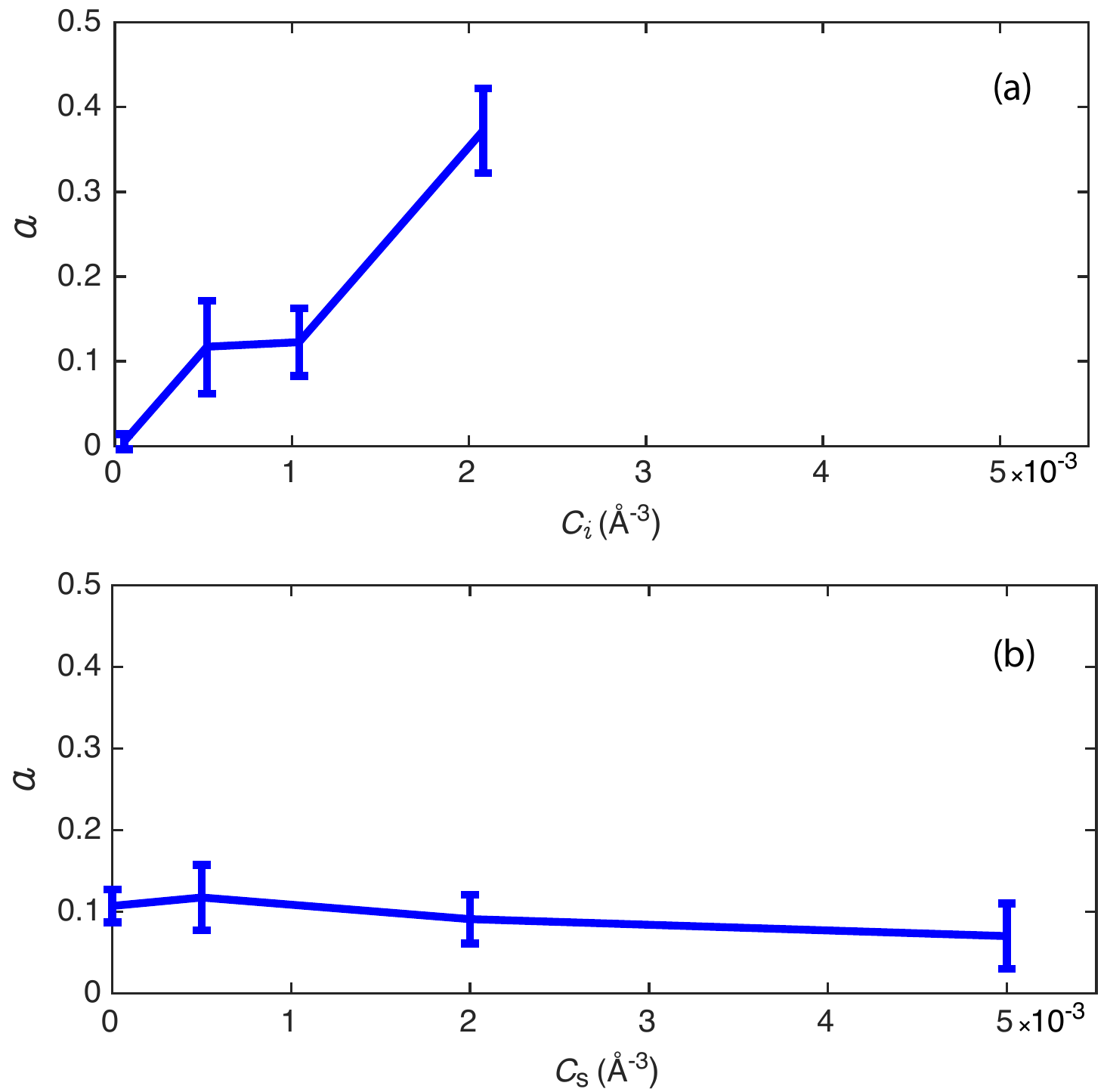}
\caption{
(a) The exponent $a$ of the fits of Fig. \ref{fig:ion-concentration} as a function of ion concentration with fixed solvent concentration, that shows a sharp increase with ion concentration.
(b) The exponent $a$ of the fits of Fig. \ref{fig:solvent-concentration} as a function of solvent concentration, with fixed ion concentration.
}
\label{fig:exponent}
\end{figure}

At low frequency, we fit the exponent $a$ of the power law $S(\omega)\sim \omega^{-a}$ for the curves shown in Figs. \ref{fig:solvent-concentration} and \ref{fig:ion-concentration}.
We fit the noise spectra for $\log_{10} \omega < -1.8$ and discard the lowest frequency data points because of their statistical uncertainty.
The exponent is shown in Fig. \ref{fig:exponent} as a function of ion concentration $C_i$ for fixed solvent concentration (top panel) and as a function of solvent concentration $C_s$ at fixed ion concentration (bottom panel).
Whereas the exponent increases sharply as a function of the ion density, increasing the solvent concentration has no effect.
Because the charge is the only difference between an ion and a solvent particle, we conclude that electrostatic interactions cause the increasing exponent.
Hydrodynamic interactions, despite having a similar long-ranged spatial dependence, do not have the same effect.

\begin{figure}[tb]
\includegraphics[width=0.9\columnwidth]{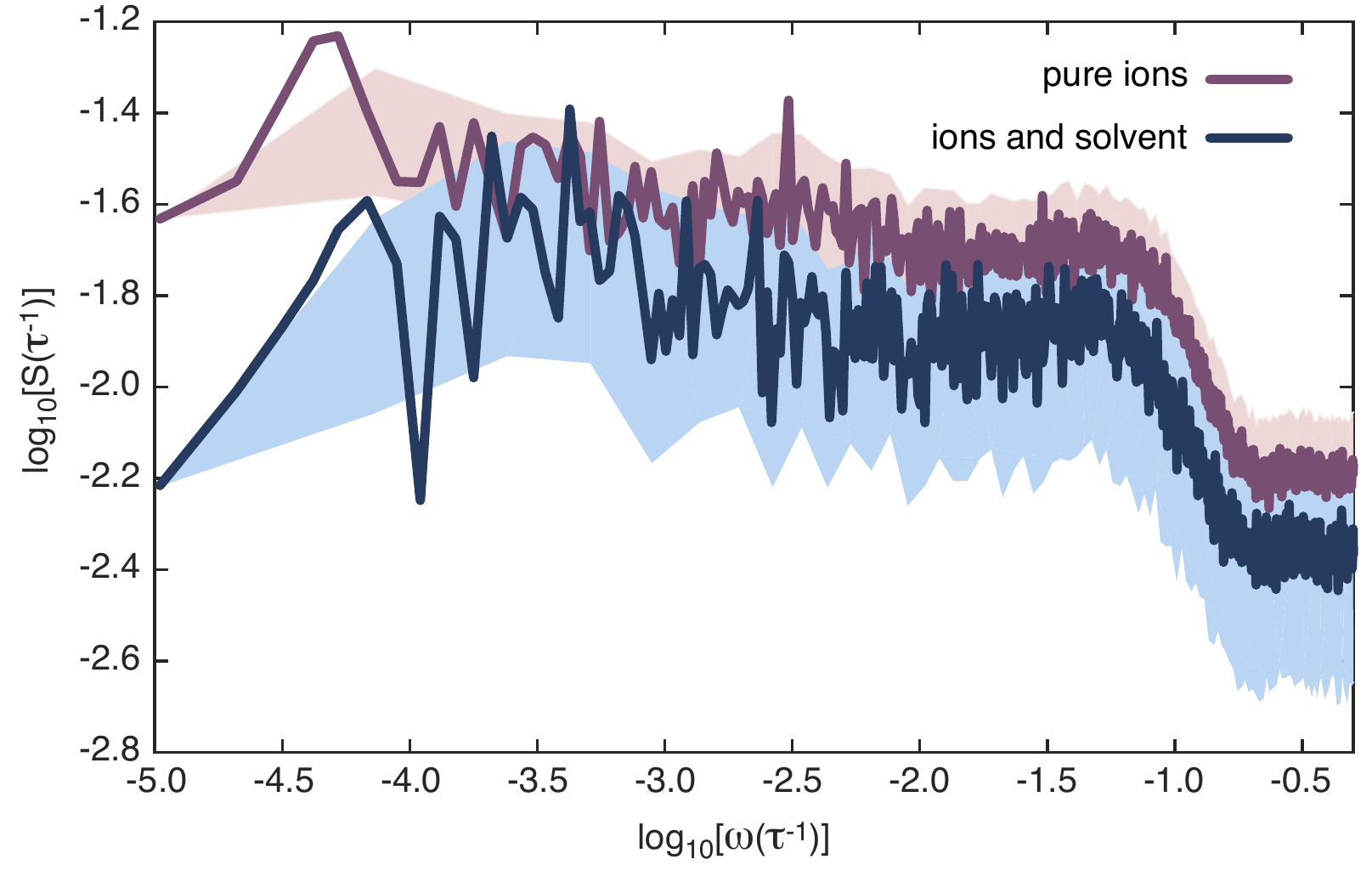}
\caption{
The power spectral density $S(\omega)$ of the current through a pore ($R=25$~\AA) filled with only ions ($C_i = 5.5 \times 10^{-4}$~\AA$^{-3}$, implicit solvent) and with ions and explicit solvent ($C_i = C_s = 5.5 \times 10^{-4}$~\AA$^{-3}$).
The electric field is set to $E_{\parallel} = 1.6$~$k_B T/(e\text{\AA})$.
}
\label{fig:head-to-head}
\end{figure}

We perform an extra simulation without solvent particles ($C_s = 0$), and compare the power spectra of simulations with and without explicit solvent directly in Fig. \ref{fig:head-to-head}.
Clearly, the curves have the same frequency dependence over the entire frequency range, confirming the results of the preceding sections.

\begin{figure}[tb]
\includegraphics[width=0.9\columnwidth]{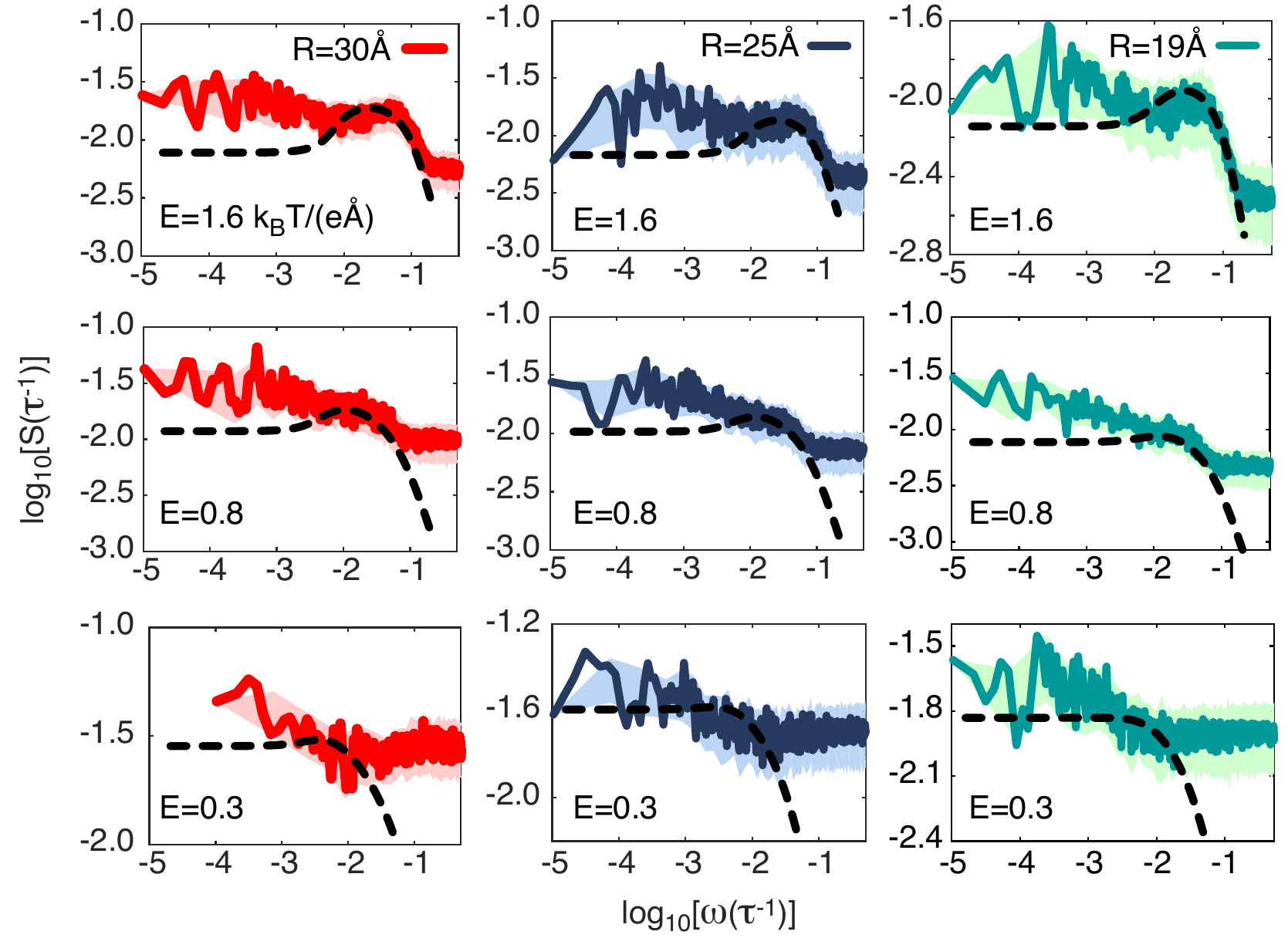}
\caption{
The power spectra of the ion current through pores with different radii $R = 19,25,30$~\AA~at equal solvent and ion concentrations of {$C_i = C_s = 5.5 \times 10^{-4}$~\AA$^{-3}$}, at different applied electric field strengths $E_{\parallel} = 0.3,0.8,1.6$~$k_BT/\left(e\text{\AA}\right)$.
With increasing $E_{\parallel}$, the transition frequency shifts upward, and the background white noise decreases. 
With increasing radius the slope at high frequency increases without any alteration of the power law at low frequency.
The dashed lines indicate the theory fit (Eq. \ref{eqn:psd}).
}
\label{fig:R-and-E}
\end{figure}

Finally, we study the dependence of the power spectral density on the pore radius $R$ and the applied electric field $E_{\parallel}$.
In Fig. \ref{fig:R-and-E}, we show that the linearized mean-field theory -- derived for implicit solvent -- captures the dependence on the pore radius and the electric field without further fit parameters,
for all values of $R$ and $E_{\parallel}$ studied.

\subsubsection{Summary and conclusions}

We present a systematic numerical investigation of the effects of hydrodynamic and electrostatic interactions on the power spectral density of ionic currents in nanopores using an explicit coarse-grained solvent.
We find that an increase in ion concentration at fixed solvent density leads to a power-law behavior at low frequency with an exponent increasing with ion density.
The power-law frequency-dependence of the power spectrum is in line with our previous findings in simulations with implicit water, where the nonzero exponent was shown to be caused by ion-ion correlations.
The exponent reaches $a = 0.4$ at an ion concentration of $C_i = 2\cdot10^{-3}$~\AA$^{-3}$.
Hydrodynamic interactions influence the power spectral density at high frequency.
In particular, the transition in the power spectral density becomes less pronounced with increasing solvent density.
At low frequency, however, the hydrodynamic interactions have no effect, which is surprising in the view of the large influence of hydrodynamic interactions on the dynamics of colloids and polymers under confinement.
Note, however, that the solvent used in the present study has a higher compressibility than water, and that the Langevin noise provides a truncation distance, which might influence the hydrodynamic interactions.
The linearized mean-field theory without hydrodynamics, which has been derived in our previous work \cite{2016_Zorkot}, can be used to describe simulation results with hydrodynamic interactions equally well.
Instead, inclusion of electrostatic ion-ion correlations is paramount to describe the low-frequency power-law behavior as a function of ion density.
Although a direct comparison with experimental results is not yet feasible, we show that the combination of simulations and analytical work provides a promising framework for the systematic investigation of experimental noise spectra.

D.J.B. acknowledges funding from the Glasstone Benefaction and Linacre College, Oxford. R.G. would like to acknowledge support from HFSP (RGP0061/2013).


%

\end{document}